\documentclass[a4paper,12pt]{article}
\pdfoutput=1
\usepackage{cite}
\usepackage{amsmath}
\usepackage{amsfonts}
\usepackage{amssymb}
\usepackage{graphicx, rotating}
\usepackage{epsfig}
\usepackage{latexsym}
\usepackage{graphicx}
\usepackage{color}
\usepackage{amsmath,bm,amssymb}

\usepackage{color}
\usepackage[english]{babel}
\usepackage{hyperref}

\newcommand{\Slash}[1]{{\ooalign{\hfil#1\hfil\crcr\raise.167ex\hbox{/}}}}

\newcommand{\beq}{\begin{equation}}  \newcommand{\eeq}{\end{equation}}
\newcommand{\bef}{\begin{figure}}  \newcommand{\eef}{\end{figure}}
\newcommand{\bec}{\begin{center}}  \newcommand{\eec}{\end{center}}
  
\newcommand{\laq}[1]{\label{eq:#1}}  

\newcommand{\Eq}[1]{Eq.~(\ref{eq:#1})}
\newcommand{\Eqs}[1]{Eqs.~(\ref{eq:#1})}
\newcommand{\eq}[1]{(\ref{eq:#1})}

\newcommand{\vev}[1]{ \left\langle {#1} \right\rangle }

\newcommand{\SU}[1]{{\rm SU{#1} } }

\def\({\left(}
\def\){\right)}
\def\dt{{d \o dt}}

\def\O{\mathcal{O}}
\def\U{\mathop{\rm U}}

\newcommand{\OR}{~{\rm or}~}
\newcommand{\AND}{~{\rm and}~}

\newcommand{\KEV}{ {\rm \, keV} }

\newcommand{\GEV}{ {\rm \, GeV} }
\newcommand{\TEV}{ {\rm \, TeV} }
\def\o{\over}
\def\a{\alpha}

\def\d{\delta}

\def\f{\phi}

\def\l{\lambda}
\def\m{\mu}
\def\n{\nu}

\def\s{\sigma}

\def\D{\Delta}
\def\G{\Gamma}
\def\L{\Lambda}
\def\F{\Phi}

\def\tl{\tilde}
\def\*{\dagger}

\begin{document}
\renewcommand\bibname{\Large References}

\begin{center}

\vspace{1.5cm}

{\Large\bf Hidden Photon and Axion Dark Matter  \\[.3em] from Symmetry Breaking}\\
\vspace{1.5cm}

{\bf  Kazunori Nakayama$^{1,2}$} and {\bf Wen Yin$^{1,3}$}

\vspace{12pt}
\vspace{1.5cm}
{\em 

$^{1}${ Department of Physics, Faculty of Science, The University of Tokyo,  \\ 
Bunkyo-ku, Tokyo 113-0033, Japan} \vspace{5pt} \\

$^{2}$Kavli Institute for the Physics and Mathematics of the Universe (WPI),
University of Tokyo, Kashiwa 277--8583, Japan \vspace{5pt}\\
$^{3}${Department of Physics, Tohoku University, Sendai, Miyagi 980-8578, Japan } 
}

\vspace{1.5cm}
\abstract{
A light hidden photon or axion-like particle is a good dark matter candidate and they are often associated with the spontaneous breaking of dark global or gauged U(1) symmetry.
We consider the dark Higgs dynamics around the phase transition in detail taking account of the portal coupling between the dark Higgs and the Standard Model Higgs as well as various thermal effects.
We show that the (would-be) Nambu-Goldstone bosons are efficiently produced via a parametric resonance with the resonance parameter $q\sim 1$ at the hidden symmetry breaking. 
In the simplest setup, which predicts a second order phase transition, this can explain the dark matter abundance for the axion or hidden photon as light as sub eV. 
Even lighter mass, as predicted by the QCD axion model, can be consistent with dark matter abundance in the case of first order phase transition, in which case the gravitational wave signals may be detectable by future experiments such as LISA and DECIGO.
}

\end{center}
\clearpage

\setcounter{page}{1}
\setcounter{footnote}{0}

\tableofcontents

\section{Introduction}
Spontaneous symmetry breaking usually happens in the thermal history of the Universe. 
Within the standard cosmology, $\L$CDM model, there are electroweak symmetry breaking (EWSB) and chiral symmetry breaking. The electroweak and chiral symmetries are restored at a high temperature due to the thermal effect but breaks when the temperature sufficiently redshift due to the expansion of the Universe. 
After the symmetry breaking, the standard model (SM) weak bosons and pions naturally appear with non-vanishing masses. 

A clear evidence of new physics beyond the SM is the presence of dark matter (DM), the origin of which is a mystery of particle theory and cosmology. Except for its longevity, abundance, and coldness, most of properties, such as the mass, spin, interactions, are not known. 
The mass of the DM may be so small and the interaction between the DM and SM particles is likely to be so weak that it is consistent with the longevity and the non-detections in various experiments e.g.~\cite{Agnese:2015nto,Akerib:2016vxi,Tan:2016zwf,Angloher:2015ewa,Amole:2017dex}. 
A simple possibility to realize both the lightness and weakness is that the DM is associated with a symmetry breaking at a high energy scale. 
Such setups are naturally realized if the DM is an axion, axion-like particle (ALP) or hidden photon similar to the pion or weak bosons (see reviews \cite{Jaeckel:2010ni,Ringwald:2012hr, Kawasaki:2013ae, Graham:2015ouw,Marsh:2015xka, Irastorza:2018dyq, DiLuzio:2020wdo}).
Then, the interaction rates are suppressed by positive powers of the mass to the symmetry breaking scale. The question is how to produce these DM candidates in the early Universe. 

A light, or explicitly a sub-keV, axion or hidden photon DM cannot be produced through thermal scatterings like the WIMP case, since otherwise it is too hot. Therefore some nonthermal production mechanism is required. It has been discussed that production mechanisms include the misalignment production for the axion~\cite{Preskill:1982cy,Abbott:1982af,Dine:1982ah} and hidden photon~\cite{Nelson:2011sf,Arias:2012az,Nakayama:2019rhg,Nakayama:2020rka}\footnote{ The most misalignment production mechanisms of the hidden photon suffer from theoretical inconsistency or observational constraints~\cite{Nakayama:2020rka}.}, gravitational or inflationary particle production for the axion~\cite{Graham:2018jyp,Guth:2018hsa} or hidden photon~\cite{Graham:2015rva,Ema:2019yrd,Ahmed:2020fhc,Kolb:2020fwh} and decay or interaction with some other fields~\cite{Randall:2015xza, Mazumdar:2015pta, Agrawal:2017eqm, Kitajima:2017peg, Agrawal:2018vin, Co:2018lka, Dror:2018pdh, Daido:2017wwb, Daido:2017tbr, Kaneta:2019zgw, Takahashi:2020uio, Moroi:2020has, Moroi:2020bkq, Abe:2020ldj}.

In this paper, we discuss the possibility that a light axion (ALP) or hidden photon DM is produced via a hidden symmetry breaking.
The simplest UV completion model to give ALP or hidden photon DM a mass is to introduce a dark Higgs field. The dark Higgs field is assumed to spontaneously break (approximate) global symmetry in the case of ALP and gauge symmetry in the case of hidden photon.
This dark Higgs field is interacting with the SM particles and thus in the early Universe the hidden symmetry is restored due to the thermal effect. The symmetry breaking occurs when the temperature becomes small enough. 
We point out that the symmetry breaking is necessary followed by a parametric resonance production of the (would-be) Nambu-Goldstone (NG) bosons if the dark Higgs is not thermalized  at the moment. In particular, we focus on a minimal setup where the dark sector and the SM sector is communicated only through the portal coupling between the SM Higgs and dark Higgs.
A large fraction of the energy of the cold dark Higgs condensate is transferred into that of the NG bosons. Consequently, the cold DM abundance is explained.\footnote{
Ref.~\cite{Dror:2018pdh} also considered hidden photon production from parametric resonance effect induced by the dark Higgs dynamics. While Ref.~\cite{Dror:2018pdh} mainly focused on the regime of broad resonance, which corresponds to the case of large initial value of the dark Higgs field, we consider the small field regime corresponding to thermal phase transition. We will show that it generically leads to marginally broad or narrow resonance and also we take into account effects of the Higgs portal coupling. }

A relevant topic may be the DM production from topological defects (see Refs.\,\cite{Yamaguchi:1998iv, Sikivie:2006ni, Kawasaki:2018bzv,Gorghetto:2018myk, Chway:2019kft} for ALPs and Ref.\,\cite{Long:2019lwl} for hidden photons.)
These defects may appear via a symmetry breaking. However, depending on the symmetry group or the breaking patterns, the defects may not appear like the case of the EWSB. In this case, our mechanism is more important.  
Even in a hidden $\U(1)$ symmetry breaking, which will be our concrete example, and which are also studied in the context of the topological defects, our mechanism provides complementary parameter regions. 
Other than the ALP or hidden photon DM, heavy DM production is discussed relevant to the bubble wall dynamics in a first order phase transition (PT) by coupling the DM to certain Higgs fields~\cite{Cohen:2008nb, Falkowski:2012fb, Baker:2016xzo, Baker:2019ndr, Azatov:2021ifm}. Compared with those studies, our DM is much lighter than the symmetry breaking scale, and our mechanism works 
not very relevant to the bubble dynamics. 
In particular, our mechanism also works in a 2nd order PT or cross-over, where the bubbles are not created. 

This paper is organized as follows. In the next section \ref{sec:2} we will discuss the NG boson production at the symmetry breaking  with  a simple Higgs portal potential, which leads to a second order PT. 
In section \ref{sec:3} we show how the DM mass is generated and how the abundance can be explained via our mechanism. 
In section \ref{sec:4} we discuss the case for a 1st order PT and the gravitational wave. 
The last section~\ref{sec:dis} is devoted to conclusions and discussion.


\section{NG boson production at symmetry breaking} 

\label{sec:2}
Let us consider the the spontaneous symmetry breaking of a hidden global continuous symmetry in the early Universe. 
Later we will gauge or explicitly break this group to give mass to the (would-be) NG boson.
In this part, we show that the NG boson can be efficiently produced soon after the symmetry breaking or PT if the dark Higgs is not thermalized at the moment. 

\subsection{Zero-temperature potential of dark and SM Higgs fields} 

To be concrete, we consider a minimal dark sector in which there is one dark Higgs field which spontaneously breaks the hidden global $\U(1)$ symmetry. 
In this minimal setup, the only renormalizable interaction between the SM and dark sector is the portal coupling between the SM and dark Higgs fields.
The most general dark and SM Higgs potential is given by
\beq
V=\L^4-\frac{m^2_\F}{4} |\F|^2+\frac{\lambda}{2} |\F|^4 +\l_P |H|^2 ( |\F|^2-v_\F^2) + \lambda_H |H|^4-\mu_H^2 |H|^2. \label{V}
\eeq 
Here $\F$ ($H$) is the hidden (SM) Higgs field (doublet) which will break the $\U(1)$ ($\SU(2)_L\times \U(1)_Y$) symmetry,  
$\lambda_P (>0), \l (>0) \AND \l_H (>0)$ are coupling constants,  $\m_H^2 \simeq (125\GEV)^2/2$ is the bare Higgs mass term in the SM, and $m^2_\F (>0)$ is the dark Higgs mass squared parameter. 
$\L^4$ is needed to cancel the cosmological constant. 
Here \beq 
v_\F \approx \sqrt{\frac{m_\F^2}{4\lambda }},
\eeq
is the dark Higgs vacuum expectation value (VEV) by introducing which in the last term we have cancelled the contribution to the SM Higgs boson mass. We will discuss the tuning to the SM Higgs boson mass, later.
The interaction between the dark sector and the SM sector is controlled by the portal coupling constant, $\lambda_P$.

Note that the portal coupling in (\ref{V}) may ensure the absolute stability of the electroweak vacuum~\cite{Lebedev:2012zw,EliasMiro:2012ay}. Assuming that $\Phi$ is much heavier than the electroweak scale, we can integrate out it below the scale $\sim m_\Phi$ to obtain the effective four-point coupling constant of the SM Higgs as $\lambda_{\rm eff}\simeq \lambda_H-\lambda_P^2/(2\lambda)$. If $m_\Phi \lesssim 10^{10}\,{\rm GeV}$ and the following condition
\begin{align}
	\frac{\lambda_P^2}{\lambda} > \mathcal O(0.01),  \label{STC}
\end{align}
is satisfied, it is shown that the quantum-corrected effective potential never becomes negative. 
If the condition (\ref{STC}) is not satisfied, either low-scale inflation or high-scale inflation with some additional SM Higgs interaction is required in order to avoid the collapse of the vaccum~\cite{Herranen:2014cua,Herranen:2015ima,Ema:2016kpf,Kohri:2016wof,Ema:2017loe,Ema:2017rkk,Figueroa:2017slm}.
On the other hand, if the condition (\ref{STC}) is satisfied, we need not to worry about such details of the inflaton and Higgs dynamics.

\subsection{Phase transition and dynamics of dark Higgs} 

In the early epoch, the Universe is filled by hot and dense plasma. 
Here we assume that the reheating occurs in the SM sector and thus the 
dense plasma, characterized by the temperature $T$, is composed by the SM particles. 
We assume that $\F$ is not fully thermalized before the symmetry breaking, i.e. 
\beq
\laq{nonth}
\G_{\rm th}^\F \ll H_{\rm ubble}\equiv \sqrt{ \frac{ g_\star \pi^2 T^4}{90 M_{\rm pl}^2}},
\eeq
with $g_{\star}$ being the effective relativistic degrees of freedom, and $M_{\rm pl}\approx 2.4\times 10^{18}\GEV$ being the reduced Planck mass. 
Here the thermalization rate of the dark Higgs field is given by 
\beq
\G_{\rm th}^\F \sim \frac{\l_P^2 T}{4\pi^3 }.
\eeq
We will show that this condition before the PT is important for our DM production mechanism to work.

At high temperature, the radial component, $s=\sqrt{2}|\F|$, gets a thermally corrected effective potential as~\cite{Dolan:1973qd,Quiros:1999jp}
\beq
V_T(s)= \lambda \frac{s^4}{8}  +
\left(\frac{\lambda_P}{6} T^2-\frac{m_\F^2}{4}\right) \frac{s^2}{2} +\cdots 
\eeq
where $\cdots$ represents irrelevant terms, including the Coleman-Weinberg corrections as well as the higher order terms.\footnote{
	Here it is assumed that $\lambda_P \gtrsim \lambda$. As will be explained later, we will mainly consider the phenomenologically preferred case of $\lambda_P^2\sim \lambda$. Then, as far as both $\lambda_P$ and $\lambda$ are smaller than unity, this assumption is justified.
}
Here is one remark. In this case we do not have a cubic term of $s$ from the $\sim -m_H(s,T)^3 T$ term in the free energy density. This is because the SM Higgs mass is $m_H^2\sim T^2 +\lambda_P (-v_\F^2+s^2/2)+2\mu_H^2$, the first term of which comes from the daisy
resummation at $s\sim 0$ \cite{Dolan:1973qd}.
Due to the contribution $-\lambda_P v_\F^2$, which is required to cancel the SM Higgs mass at the vacuum $s\sim \sqrt{2}v_\Phi$,
$m_H^2$ cannot be approximated as $m_H^2\sim \lambda_P s^2$ for $s \lesssim \sqrt{2}v_\F$.
Thus there is no parameter region for $s^3$ term to appear. 
This is a peculiar feature of the SM Higgs contribution to the thermally corrected potential: the SM Higgs is (almost) massless at the finite VEV of $s=\sqrt{2}v_\Phi.$
Thus we expect that the PT of $\Phi$ is the second order.

As the Universe expands, the temperature $T$ decreases. One can easily see that the symmetry is broken at the temperature
\beq
T\lesssim T_{\rm crit}=\sqrt{\frac{3m_\F^2}{2\lambda_P}}.
\eeq
As we have explained above, there is no cubic term or potential barrier in the thermal effective potential and hence the PT is expected to be the 2nd order. 
A similar discussion can be also made to the SM Higgs potential, which is broken while $\U(1)$ is symmetric if $T^2 \lesssim \max{[\lambda_P v_\F^2,  \mu_H^2]}$.
We can easily find that the symmetry breaking of $\U(1)$ occurs prior to the electroweak symmetry breaking if 
\beq
\laq{cond1}
\lambda \gtrsim \max{\left[\lambda^2_P,\frac{\lambda_P \mu_H^2}{v_\F^2}\right]}.
\eeq
This condition will be assumed, so that we can safely neglect the dynamics in the $H$ direction.

Now let us see the $s$ dynamics around the PT.
When $T\lesssim T_{\rm crit}$, the VEV of $\F$ is temperature dependent,
\beq
v_\F^T= \sqrt{\frac{1}{\lambda}} \frac{m_\F^{T}}{2},
\eeq
with 
\beq
\laq{MT}
m_\F^{T}= \sqrt{m_\F^2-\frac{2\lambda_P T^2}{3}},
\eeq
being the temperature dependent effective mass. 
The radial component of the dark Higgs $s$ may follow the potential minimum just at around the transition, but after the transition $s$ starts to oscillate around the temperature dependent minimum if 
\beq \laq{osc}  m_\F\gg H_{\rm ubble}. \eeq
The energy density of the coherent oscillation comes from part of the potential energy~$\L^4$.
This can be seen by solving the equation of motion in a simplified setup by neglecting the contribution from the NG bosons:
\beq
\ddot{s}+3H_{\rm ubble}\dot{s}= -\frac{\partial}{\partial s}V_T. 
\eeq
We will solve this equation by taking the initial conditions $s(0)=0$, 
and \beq 
\laq{fluc}
\dot{s}(0)=\frac{\G_{\rm th}^\Phi}{H_{\rm ubble}} T_{\rm crit}^2.
\eeq 
We set $\dot{s}$ with a tiny non-vanishing value initially, because we expect that there is a thermal fluctuation, which kicks $s$ at random. 
If $\F$ is initially thermalized, we expect $\dot{s}\sim T^2$. We take into account of the thermal fluctuation by this initial condition. As we will see soon that this initial condition is insensitive to the result as long as \eq{nonth} is satisfied.

\begin{figure}[!t]
\begin{center}  
   \includegraphics[width=75mm]{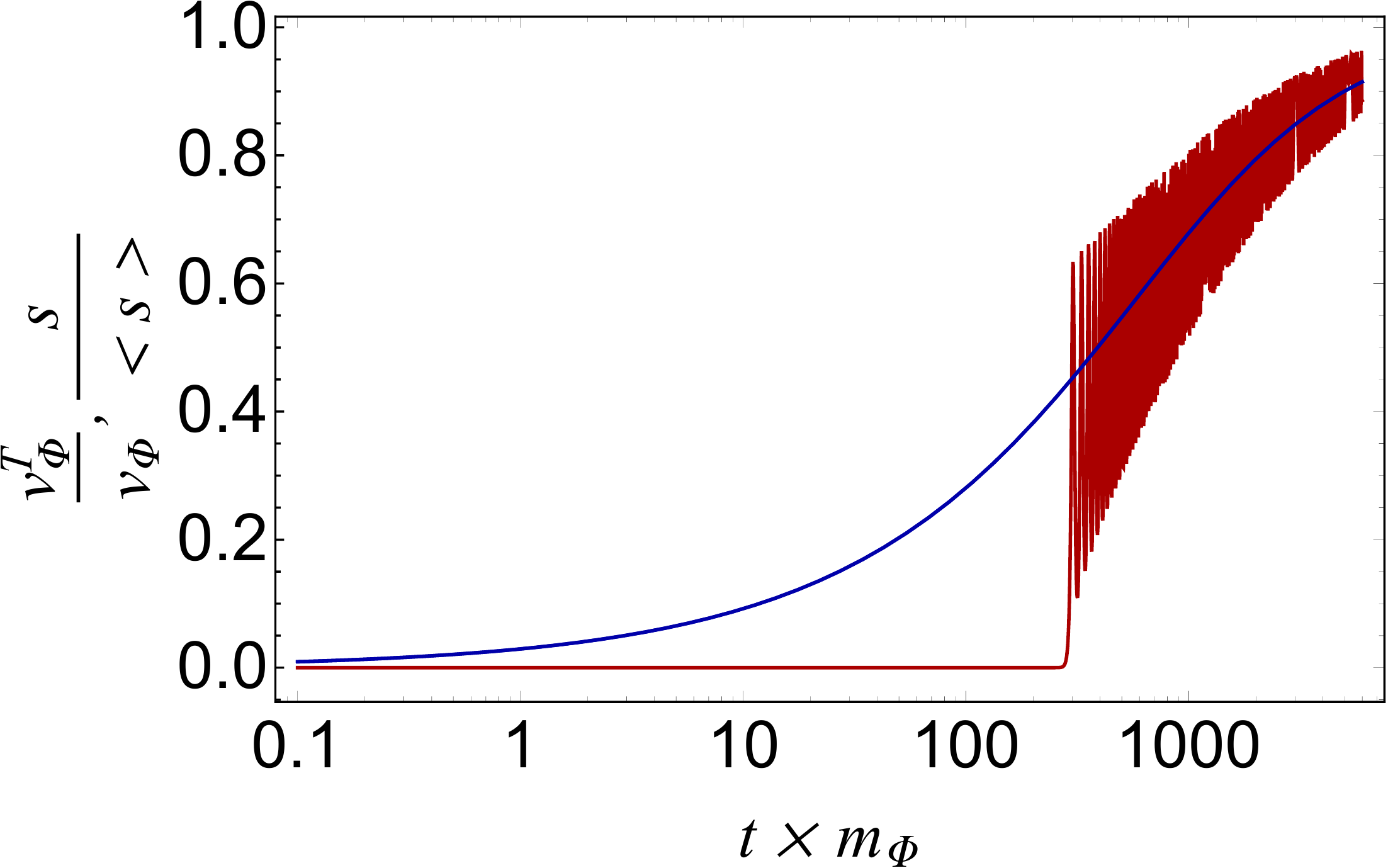}
      \includegraphics[width=78mm]{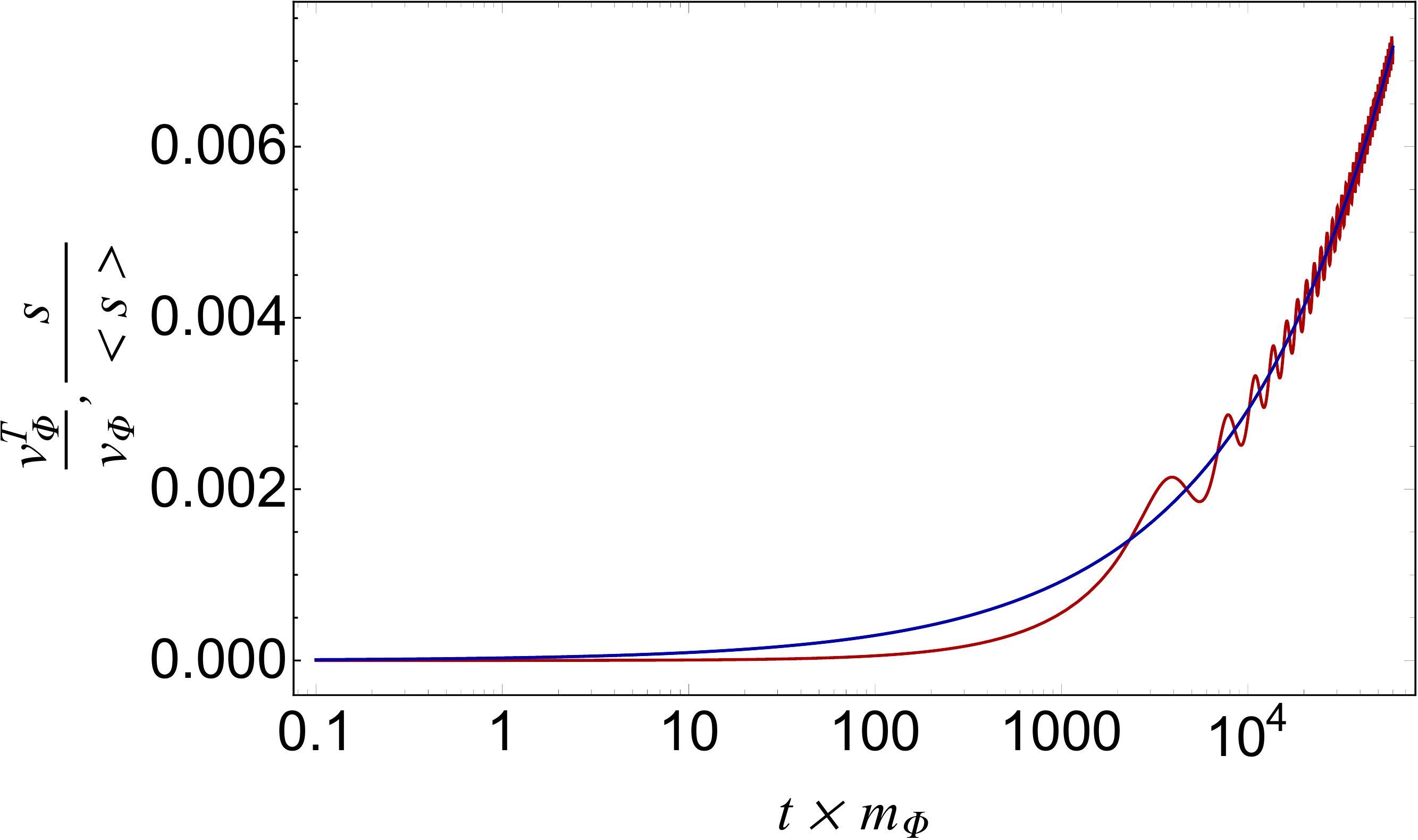}
      \end{center}
\caption{
The time dependence of $s$ and $v_\F^T$ around the 2nd order PT given in red and blue solid lines, respectively.
Effects of particle production and dissipation/decay are neglected. In the left panel we have taken $m_\F=0.1\GEV, v_\F=10^{14}\GEV$, and in the right panel $m_\F=0.1\GEV, v_\F=10^{8}\GEV$. 
We fix $\l=\l_P^2.$  } \label{fig:osc}
\end{figure}

In Fig.~\ref{fig:osc}, we show the numerical result of $s/(\sqrt{2}v_\F)$ [red solid line] and $v_\F^{T}$ [blue solid line] with $\l=\l_P^2$. 
In the left panel, where $m_\F=0.1\GEV, v_\F=10^{14}\GEV$, the oscillation takes place within a few $1/m_\F$ which is much shorter than one Hubble time ($m_\F/H_{\rm ubble}(T_{\rm crit})\simeq 2338$ in this case). 
After the onset of oscillation, the oscillating amplitude decreases in time. 
We also notice that $v_\F^{T}$ is settled into $v_\F$ 
within $\O(1)$ Hubble time. 
This is the case $\G_{\rm th}^\F/H_{\rm ubble}\simeq 2.58\times 10^{-22}.$
 When $m_\F\gg H_{\rm ubble}, \AND \G_{\rm th}^{\F}/H_{\rm ubble}$ are larger, the transition is faster, as shown in the right panel.  
 Here   $m_\F=0.1\GEV, v_\F=10^{8}\GEV$  and $ \G_{\rm th}^{\F}/H_{\rm ubble}\simeq 2.58\times 10^{-7}$. In this case the oscillation amplitude is much smaller

To understand this behavior, let us consider two time periods for the $s$ evolution. 
Soon after the PT, $s$ is placed at the hilltop with a non-vanishing negative mass squared. Then $s$ starts to slow-roll towards the bottom of the potential as long as there is a tiny $\dot{s}$ or $s$ at $T\simeq T_{\rm crit}.$  
The slow-roll lasts much longer than $1/m_\F$ because of the vanishingly small $V_T'$ at around the hilltop. 
The time, $\D t_{\rm osc}$, for the onset of the oscillation (measured from the instant of $T=T_{\rm crit}$) can be obtained by solving the equation of motion with neglecting the Hubble friction. 
We can take $s \propto \exp{\left(\int_0^t {m_\F^T d t'}\right)}$. 
When the exponent becomes larger than $\O(1)$, the slope becomes so steep and evetually $s$ starts to oscillate. Since the evolution is in exponential, the result will not be very sensitive to the initial condition. 
By expanding $m_\F^T$ around $T\sim T_{\rm crit}$, we obtain the time scale for the onset of oscillation \beq \laq{Dtosc}\D t_{\rm osc}\sim \tl{C} (m_\F^2 H)^{-1/3},\eeq after the PT.  
 The coefficient, $\tl{C}$, which is not very different from $\O(1)$, logarithmically depends on the initial condition \eq{fluc} when it is small enough. After the onset of oscillation, the number density is an adiabatic invariant, 
\beq
n_s^{\rm osc} \approx \left.\frac{m_\F^T}{2} (v_{\F}^T)^2\right|_{\text{onset of oscillation}}. 
\eeq
Then we can define 
\beq
C_{\rm eff}\equiv \frac{2n^{\rm osc}_s}{m_\F v_\F^2},
\label{Ceff}
\eeq
for later convenience. It represents an effective suppression factor of the $s$ coherent oscillation abundance.
It is evaluated as $C_{\rm eff}\sim 10 \tl{C}^{3/2} m_\F/( M_{\rm pl} \l_P)$ by inserting $v_\F^T$ at the timing of onset of $s$ oscillation \eq{Dtosc}.

\begin{figure}[!t]
\begin{center}  
   \includegraphics[width=105mm]{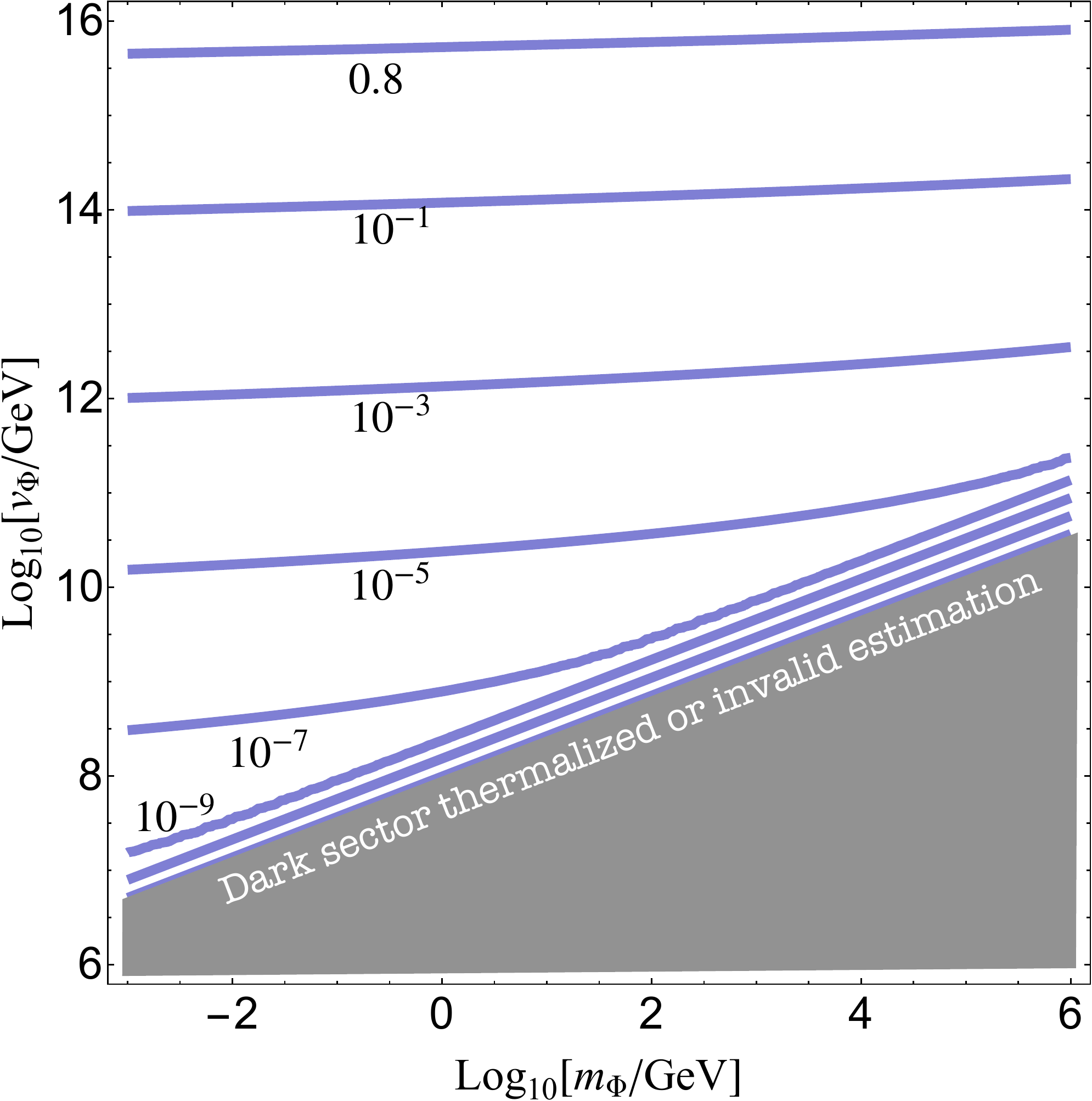}
      \end{center}
\caption{
The contours for $C_{\rm eff}\equiv \frac{2n^{\rm osc}_s}{m_\F v_\F^2},$ on $(m_\Phi, v_\F)$ plane for $\l_P^2=\l$. 
 In the grey shaded region the dark Higgs is (close to be) thermalized at the phase transition or our estimation here is invalid.} \label{fig:contour}
\end{figure}

The contours of $C_{\rm eff}$ by numerically solving the equation of motion is given in the $(m_\F, v_\F)$ plane in Fig.~\ref{fig:contour}. For simplicity we take $\l =\l_P^2$, in which case $C_{\rm eff}$ scales as $10 (\tl{C}^{3/2}v_\f/ M_{\rm pl}).$ 
We find a non-negligible number of $n_s$ produced due to the oscillation of $s$ after the PT. 
We emphasize that the discussion is based on the condition that $\F$ is not thermalized or $\G_{\rm th}^{\F}\ll H_{\rm ubble}$ at the PT.  
When $\G_{\rm th}$ and $H_{\rm ubble}$ are close to each other soon after $T= T_{\rm crit},$  
 $s$ starts to oscillate without being trapped at around the potential top. 
This will lead to a different conclusion with a much suppressed NG boson production.

Here we emphasize that we neglect the particle production of the (would-be) NG boson for illustrative purpose. 
In practice, we cannot neglect the interaction with the NG boson. As we will show soon, the NG boson production happens at a similar time scale of the oscillation frequency,  $m_\F^T$.

\subsection{Particle production at the phase transition}

Let us consider the particle production at around the onset of oscillation, i.e. at around the symmetry breaking. By writing $\Phi = v_\Phi +(s+ia)/\sqrt{2}$, the $s$ condensate decays to the SM Higgs pair via ${\cal L}\supset -\sqrt{2}\lambda_P v_\F s |H|^2 $  and the NG mode, $a$, via\footnote{ The results are the same for the non-linear parametrization $\Phi= v_\Phi e^{(s+ia)/\sqrt{2}}$ in which case the decay occurs via the kinetic coupling ${\cal L}\supset \frac{s}{v_\F} (\partial a)^2.$ } ${\cal L}\supset -\frac{\lambda}{\sqrt{2}} v_\F s a^2$. 
The decay rate to the SM Higgs multiplets is given by 
\beq
\laq{decayH}
\G_{s\to HH}\approx \frac{\lambda_P^2 v_\F^2}{4\pi m_\F}, 
\eeq
by neglecting the SM Higgs boson mass. However, when $T\gtrsim m_\F$, this process is kinematically forbidden due to the heavy thermal mass of $H$. Instead, as we shall see later, a thermal dissipation effect may work.

\paragraph{Parametric resonance production of $a$-condensate}

The decay rate to the NG mode pair, on the other hand, is given by 
\beq
\laq{decaya}
\G_{s\to aa }= 
 \frac{\l }{16 \pi }m_{\Phi}=\frac{m_\F^3}{64\pi v_\F^2 }.
\eeq
By taking into account the thermal corrections, $m_\F$ in the r.h.s should be replaced to be $m^{T}_{\F}.$ 
With oscillating $s$, we should take into account the parametric resonance effect for $s \to a$'s since there could be a Bose-enhancement effect~\cite{Traschen:1990sw, Kofman:1994rk, Shtanov:1994ce, Yoshimura:1995gc, Kasuya:1996np, Kofman:1997yn, Dufaux:2006ee, Matsumoto:2007rd, Asaka:2010kv, Amin:2019qrx, Garcia:2018wtq, Lozanov:2019jxc, Alonso-Alvarez:2019ssa, Moroi:2020has, Moroi:2020bkq}. 
 The resonance $q$-parameter~\cite{Kofman:1997yn} can be obtained for the $s$-$a$ system soon after the PT as
 \beq
 \laq{qa}
 q_a\equiv \frac{ 2 \sqrt{2}\lambda v^T_\F s_{\rm amp} }{(m_\F^T)^2}\lesssim 1,
 \eeq
where $s_{\rm amp}$ denotes the oscillation amplitude of $s$ and we have used $s_{\rm amp}\lesssim v^T_\F.$
In particular, this is maximized at the onset of oscillation, where $s_{\rm amp}\sim v^T_\F.$
During the parametric resonance the phase space distribution function of $a$ in the resonance band increases exponentially as 
  $f_{k}\propto \exp{(\frac{q_a}{2} m^T_\F t)}$ for $k\simeq m^T_\F/2$.
Thus, the growth rate $q_a m^T_\F$ is as fast as the oscillation $m^T_\F$ at the onset of oscillation. 
This is a generic prediction associated with the Higgs dynamics around the symmetry breaking in the early Universe. 
Since $m^T_\Phi$ is much larger than the Hubble scale (\Eq{osc}), the parametric resonance effect transfers the most of the $s$ oscillation energy density to the NG mode $a$ after $\O(1)$ oscillations of the $s$ field. The resulting number density of $a$ from this effect is estimated as
\beq
\laq{numd}
\frac{n_a^{(\rm para)}}{S} \sim 
 \frac{C_{\rm eff}m_\F v_\F^2}{\frac{2\pi^2 g_{\star, s }}{45} T_{\rm crit}^3},
\eeq
with $S$ being the entropy density of the SM plasma, $g_{\star, s}$ being the relativistic degrees of freedom for the entropy density and $n^{\rm (para)}_{a}$ the  produced number density of $a$. Here we have taken the temperature at the onset of oscillation to be $T_{\rm crit}$.

The parametric resonance effect for the $s$-$H$ system, on the other hand, is suppressed again due to the large thermal mass  of $H$. 
Although $s$ cannot decay into $H$, $s$ can scatter with the thermal plasma at a rate~\cite{Yokoyama:2005dv,Anisimov:2008dz, Drewes:2010pf, Mukaida:2012qn, Drewes:2013iaa, Mukaida:2012bz, Mukaida:2013xxa, Moroi:2014mqa}\footnote{We note this is different from some of the result in the references due to the time scales where we consider $m_\F\ll \G_{\rm th}, T $, with $\G_{\rm th}$ being the thermalization rate of the Higgs, $\G_{\rm th} \sim y_t^2 T$. For example, Eq.~(27) of Ref.~\cite{Yokoyama:2005dv} implicitly assumes $\G_{\rm th}\ll m_\Phi \ll T$ and it is different from our expression. On the other hand, Eq.~(3.34) of Ref.~\cite{Mukaida:2012qn} is derived under the same assumption as ours and hence the result is consistent.}
\beq
\laq{diss}
\G_{s H\to t \bar{t}}\sim  \frac{\lambda_P^2 v_\F^2 y_t^2}{4\pi (y_t^2 T^2)^2}\times  \frac{T^3}{\pi^2}\sim \frac{\lambda_P^2 v_\F^2}{4\pi^3 y_t^2 T},
\eeq
where we have considered the production of a top quark pair with $y_t\sim 1$ being the top Yukawa coupling by taking account of the Higgs thermal mass. 
Here we have neglected the temperature dependence on $v_\F$ for the time scale to consider. 
This can be also found from the equation of motion of $s$ in an effective action
(see appendix \ref{ap:1}).
Through this process, the energy stored in $s$ may be dissipated into the SM plasma. 
In order for the parametric resonance effect not to be blocked by such a scattering process, we need~\cite{Kofman:1997yn }
\beq
\laq{cond}
q_a m_\F^{(T)} \gtrsim \G_{ s H\to t\bar{t}}.
\eeq
This is easily satisfied soon after the onset of oscillation (see \Eqs{cond1} and \eq{MT}). 
Consequently, soon after the PT, $s$ has a good environment for producing $a$ via parametric resonance.

\paragraph{Dissipation of $a$-condensate }

We also need to consider the dissipation of produced $a$ soon after the PT. The dissipation rate of the $k$-mode of the NG boson is found to be (see the Appendix~\ref{sec:dissipation} for derivations):
\beq
\laq{dis}
\G^{(a)}_{\rm  dis}[n_{a}] \sim \frac{\lambda_P^2 k^2 n_{a}}{8\pi^2 m_\F^4} \frac{k}{y_t^2 T}.
\eeq
This can be seen as the $a\text{-}a$ annihilation and thus it is proportional to the number density of the $k$-mode NG boson. 
This effect is most important at the NG boson production $k \sim m_\F/2 \sim \sqrt{\lambda_P} T_{\rm crit}/2$.\footnote{Strictly speaking, $k \sim C_{\rm eff}^{1/3} m_\f/2$ in the second order phase transition. This is because at the production of the NG boson, the mass of $s$ is smaller due to the aforementioned thermal correction. 
As we will see, the annihilation effect, even that it is overestimated, is not important in the second order case. 
On the other hand, in the first order PT case in Sec.\,\ref{sec:4}, the production of $a$ is delayed and the momentum is not suppressed by $C^{1/3}_{\rm eff}.$ 
Thus \Eq{dis} is more accurate. 
}
We can neglect the annihilation effect if \beq \G^{(a)}_{\rm dis}[n_a^{\rm (para)}]< H_{\rm ubble}\eeq soon after the production, 
so that $a$ is kept intact. 

If this is not satisfied, some of the NG bosons annihilate into the SM plasma, while some NG bosons remain. 
The remnant $n_{a}$ can be estimated from $\G^{(a)}_{\rm  dis}\sim H_{\rm ubble}$ as
\beq
\laq{anni}
\frac{n_{a}^{\rm (ann)}}{S}\sim 2\times 10^{-3} \sqrt{\frac{106.75}{g_{\star}}} \({\frac{T_{\rm crit}}{10^5\GEV}}\)^4\(\frac{100\GEV}{m_\F}\)^3
\eeq
like the WIMP scenario. 
Therefore soon after the PT we get 
\beq
\frac{n_{a}}{S}\approx \frac{\min{[ n_a^{\rm (ann)}, n_a^{\rm (para)}]}}{S},
\eeq
which originates from the PT. Much after the PT, the annihilation effect on $a$-condensate is suppressed and we will neglect it. \\

Before moving to the next section, let us discuss some other components produced relevant to the PT.
\paragraph{Remnant of $s$-condensate}
The efficiency of the parametric resonance for producing the NG mode $a$ decreases as the amplitude of $s$ decreases due to the energy transfer into $a$. Eventually the parametric resonance stops and a tiny component of the $s$-condensate should remain.
It is estimated as follows. The resonance peak is at $k\simeq m^{T}_\Phi/2$ while the width of the resonance band in the momentum space is given by $\sim q_a m^{T}_\F$. The redshift of the momentum and the enhancement of $m^{T}_\F$ due to the Hubble expansion takes the produced $a$ away from the resonance band within a time scale $\Delta t\sim  q_a/H_{\rm ubble}$. The resonant enhancement stops when the exponential growth factor $q_am_\F \Delta t\sim q_a^2 m_\F/H_{\rm ubble} \sim 1$.
From this we can estimate $s_{\rm amp}$ at which the parametric resonance stops and hence the remnant of the energy density of $s$ as \cite{Kofman:1997yn, Moroi:2020bkq}\footnote{As far as the narrow resonance with $q_a\ll 1$ is concerned, we expect that the back-reaction such as $aa\to ss, \OR s a \to s a$ is not important because it is either kinematically invalid or the rate is suppressed by $q^4_a.$
On the other hand, due to the tachyonic instability the fluctuation of $s$ itself may also develop within a few oscillations~\cite{Felder:2000hj,Felder:2001kt}, which may tend to stop the resonant enhancement of $a$. Still, however, the conclusion that the most energy of $s$ is transferred to $a$ should remain valid. 
 The condition \eq{cond} may not be satisfied for a smaller $q_a$. In such a case, before the Hubble expansion becomes important in preventing the production of $a$, the dissipation may be more important. 
 In this case, the $s$ condensate is easier to thermalize than our estimation which does not change our conclusion. 
} 
\beq \left. m_\F n_s^{\rm rem}\equiv \rho_s\right|_{T\lesssim T_{\rm crit}}\sim C_{\rm eff}^{2/3}\frac{H_{\rm ubble} m_\F^3}{\pi \l }.\eeq  
One can also estimate the order of it from the Boltzmann equation by taking account the Bose-enhancement effect~\cite{Moroi:2020has}. 

\paragraph{Topological defects}
After the PT, topological defects may be formed. 
In our $\U(1)$ case, there are cosmic strings produced after the PT. 
Cosmological effects of cosmic strings in our scenario will be briefly discussed in the next section.

\section{Dark matter production at second order phase transition}
\label{sec:3}

Let us apply the mechanism of NG boson production discussed in Sec.\ref{sec:2} to the DM production. 
The DM, if dominant, must be cold and thus we should somehow give mass to the NG boson to make it non-relativistic around and after the galaxy formation era. 

In Secs.\,\ref{sec:32} and \ref{sec:33}, we will provide two possibilities for generating masses of the DM: explicitly breaking the global $\U(1)$ and gauging the $\U(1)$, in which case the DM becomes an axion(-like particle) and hidden photon, respectively. 
In the latter case, we can discuss the most properties of DM by looking at the NG boson Lagrangian according to the equivalence theorem since we are interested in the light hidden photon DM and it is highly relativistic at the production. 
Thus, in the Sec. \ref{sec:31}, we first discuss general model-independent features by assuming that the NG boson acquires a mass term of $m_{a}$ and discuss the thermal history after the PT.

\subsection{Dark components after phase transition}
\label{sec:31}

(Not much) After the PT, we have five kinds of cosmic components other than the SM particle plasma: 
the $a$ condensate from parametric resonance, the remnant $s$ condensate, the topological defects, 
$a$ produced from thermal scattering,  and $s$ produced from thermal scattering.
We use the ``condensate" to distinguish the ``cold" component, whose typical momentum is much smaller than the cosmic temperature $T$, which is the typical momentum of a ``particle" from the thermal scattering. 
They will be discussed separately.

As we will discuss soon, the remnant of $s$ may be dissipated away due to \eq{diss}, decay via \Eqs{decayH} or via the mixing with the SM Higgs boson when the Universe cools down. 
For simplicity, let us assume that the remnant of $s$ condensate does not dominate the Universe and its subsequent interactions do not play important role on cosmology.\footnote{In the case that a dominant $s$ decays at late time, we need to take account of the entropy dilution to the DM abundance or dark radiation constraints on $a$. 
The spectrum of the dark radiation of $a$ can be an evidence of the reheating if it is measured~\cite{Jaeckel:2021gah}. }  
Thus, the $s$ will neither contribute to nor dilute the DM abundance. Due to this assumption, we can first calculate the DM. 
We will check that this assumption is satisfied in the parameter region of interest. 
We will also come back to the case that $s$ once dominates the Universe in the last section, by considering $s$ as an inflaton.

\paragraph{$a$-condensate as dominant dark matter component}
The produced $a$ condensate later composes the DM when it acquires the mass $m_a$ and becomes non-relativistic. 
We can calculate the abundance of the (would-be) NG boson $a$ from
\beq
\laq{abundance}
\Omega_a  \sim  \frac{m_{a} n_a}{S}\frac{S_0}{\rho_{\rm c}},
\eeq
where $S_0~(\rho_{\rm c})$ is the entropy density (critical density) today. 
This explains the observed DM abundance if~\cite{Aghanim:2018eyx} 
\beq \Omega_a h^2 = \Omega_{\rm DM} h^2 \sim 0.12, \eeq
with $h\simeq 0.67$ being the present Hubble parameter in unit of $100\,{\rm km/s/Mpc}$.
Also, to explain the coldness of the DM we use the conservative bound calculated in \cite{Moroi:2020has} (see also Refs.~\cite{Viel:2005qj, Irsic:2017ixq}),
\beq
\laq{coldness}
C_{\rm eff}^{1/3}\frac{10^{-6}\GEV}{m_{a}} \frac{m_\F}{T_{\rm crit}}\lesssim 1,
\eeq
which gives a lower bound on the DM mass. 
Interestingly, since this becomes 
\beq
 C_{\rm eff}^{1/3}\frac{\lambda_P^2}{\lambda} \frac{0.07}{\Omega_a h^2}\lesssim 1,
\eeq 
 by using \Eq{abundance}, the coldness bound is automatically satisfied from \Eq{cond1}. 
 Thus the DM from the symmetry breaking is naturally cold. 

\begin{figure}[!t]
\begin{center}  
   \includegraphics[width=105mm]{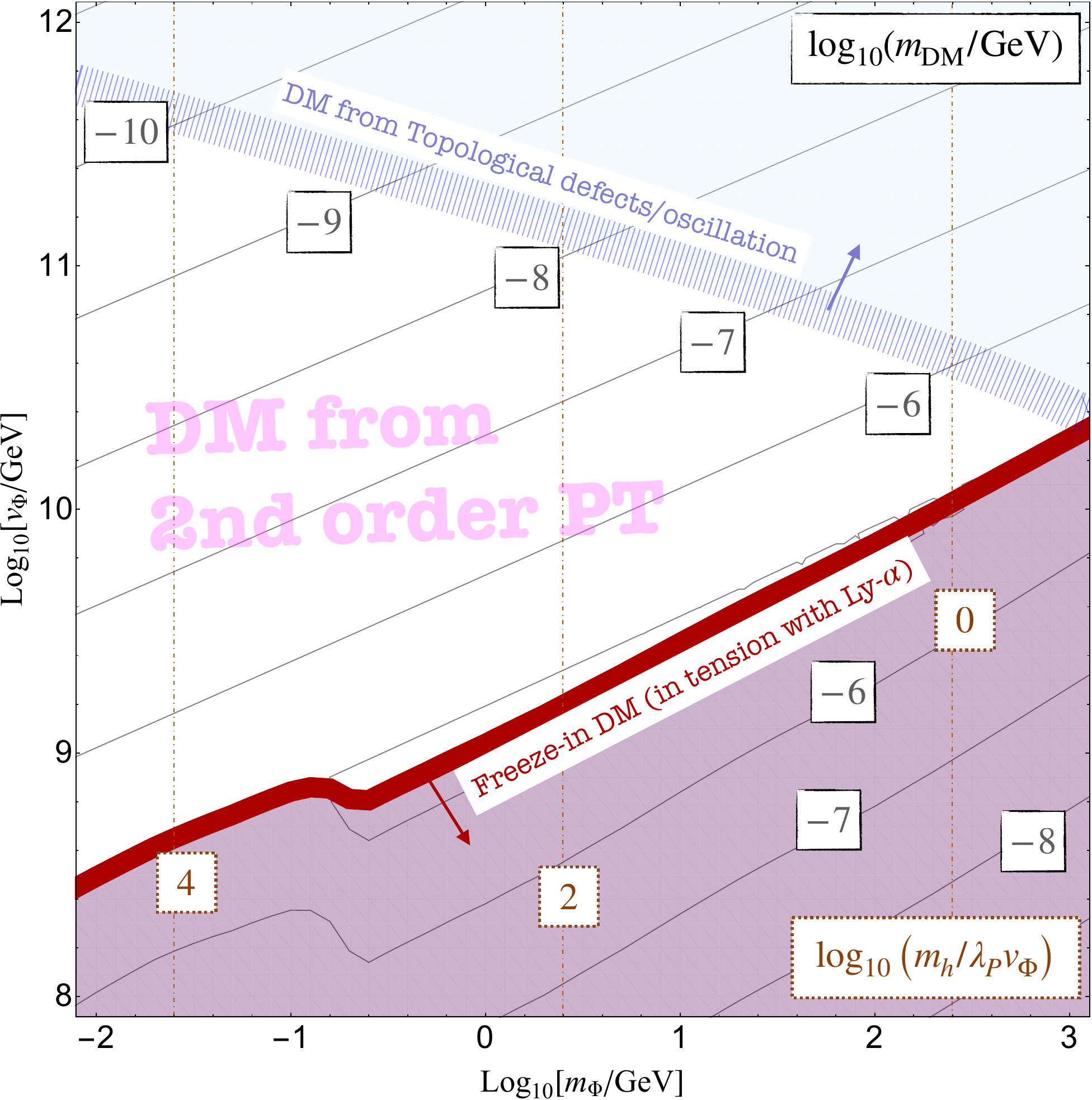}
      \end{center}
\caption{
The contours for the lightest possible DM mass in $(m_\Phi, v_\F)$ plane.  
The vertical dot-dashed contours in orange denote the amount of fine-tuning to the SM Higgs boson mass, $m_h/\l_P v_\F.$ 
Various constraints discussed in the main part are included. The DM could be either an ALP or hidden photon. 
Below the red line the DM is mostly produced from the freeze-in mechanism which is in tension with the Lyman $\a$ data. 
Above the blue dashed band the topological defect/coherent oscillation may be dominant. } \label{fig:DM}
\end{figure}

For explanation of the effects of the following constraints from the thermal history, 
 we first show the contour plot of the DM mass given in Fig.~\ref{fig:DM}. 
Again here we take \beq \laq{coupling}\lambda_P^2 = \lambda, \eeq 
which is the largest $\l_P^2$ satisfying $\eq{cond1}$, and corresponds to (almost) the lightest DM according to \eq{coldness}.
Note that this choice is consistent with the condition for the absolute stability of the electroweak vacuum (Eq.~(\ref{STC})).

\paragraph{Thermal history for the remnant $s$-condensate and $s$ particle}
 To discuss the evolution of other components, let us introduce $T_{s\to {HH}}\AND T_{s\to aa}$  which are defined by $T_{i} \equiv \(\frac{ g_{\star} \pi^2 }{30 }\)^{-1/4} \sqrt{M_{\rm pl} \G_{i}} $. 
As we have explained that the decay $s\to { HH}$ is thermally blocked and dissipation is important. $T_{s\to HH}$ should not be considered as the decay temperature. 
The dissipation rate \eq{diss} is smaller than $\G_{ s \to  HH}$ with $T\gtrsim m_\F$, 
comparable to $\G_{s\to HH}$ with $T \sim m_\F$. In fact if $T\lesssim m_\F$ the dissipation effect is suppressed since $\G_{s H \to t t} \propto (\lambda_P v_\F)^2 T/m_\F^2$ for $T\lesssim m_\F.$
Therefore the dissipation can remove $s$ condensate away if and only if \beq \laq{discon} \G_{s \to HH}\gtrsim H_{\rm ubble}[T\sim m_\F],\text{  i.e.  } T_{s\to HH}\gtrsim m_\F\eeq 
If \eq{discon} is satisfied, $T_{s \to HH }\gtrsim T_{\rm th}\gtrsim m_\F,$ where $T_{\rm th} $  is defined with $\left. \frac{\G_{s H\to tt}}{H_{\rm ubble}}\right|_{T=T_{\rm th}}= 1$.
In fact, in the figure, $T_{\rm th}$ is always greater than $m_\F $ and the electroweak scale. Thus $s$-condensate evaporates. 

 We must also consider the thermal production of $s$ particles since the production rate, which is dominated by the inverse decay, is given as
$
\G_{HH \to  s} \sim  \frac{(\lambda_P v_\F)^2}{4\pi T}.
$
The production rate via $tt \to s H$  has a similar form.
This is comparable to $\G_{s H \to tt}.$ 
Then at $T\sim T_{\rm th}$, the $s$-condensate disappears, but, instead, $s$-particles are thermalized. 
The thermalized $s$ mostly interacts with the SM particles if $T_{s\to aa}<m_\F$, until $s$ becomes non-relativistic. 
On the other hand, if $T_{s\to aa}>m_\F$, $a$ particles are produced via the $s$ decay and $a$ are also thermalized. 
In the end, $s$ would decay to SM thermal plasma. 
Since we focus $\l \sim \l_p^2$, the decay rate to $aa$ is smaller than the decay rate to $HH$ if $\l \lesssim1$.\footnote{We note that we may also consider the $s$
decay to $aa$ 
when $\l^2 \gtrsim \l_P^2$ in general.  } 
Since \Eq{discon} in the parameter region of interest, 
the components of $s$ condensate and particles disappear from the Universe not much later than $T\sim m_\F.$  
The decay of $s$ should not cause cosmological problems as long as they happen at a high enough temperature. 
In particular, we take
\beq
m_\F\gtrsim 0.01\GEV. 
\eeq
from the viewpoint of the big-bang nucleosynthesis~\cite{Kawasaki:1999na, Kawasaki:2000en,Hannestad:2004px, Ichikawa:2006vm, DeBernardis:2008zz, deSalas:2015glj, Hufnagel:2018bjp, Hasegawa:2019jsa, Kawasaki:2020qxm,Depta:2020zbh}.\footnote{Since the mass is relatively heavy, we neglect bounds on $s$ particles from stellar cooling arguments.}
This is the lower limit of the horizontal axis of the figure. 
After the decoupling/decay of $s$, the Universe is composed by three components: 
$a$-condensate, $a$-particles, and topological defects.

We also mention that $s$ in the sub GeV mass range, which mixes with the SM Higgs with a mixing angle $\theta_H \sim \lambda_P v_\F/m_h,$ $\theta_H\gtrsim  10^{-3} $, may be excluded by the accelerator bounds or BBN constraint. A large fraction of the allowed range may be tested in the SHiP experiment~\cite{Winkler:2018qyg}.

\paragraph{Freeze-in production of $a$}

Although $s$ dominantly decays into SM particles (via mixing with the Higgs if it is lighter than $2m_h$), the rare decay into $aa$ provides a freeze-in production of DM. 
The produced abundance of $a$ can be estimated as 
\beq
\Omega_a^{\rm th}\sim \frac{S_0}{\rho_c} \left.\frac{2n_s}{S} \times  \frac{\G_{s\to aa}}{H}\right|_{\rm T\sim m_\F}.
\eeq
This explains the DM abundance, $\Omega_a^{\rm th}\sim \Omega_{\rm DM}$ with $m_a$ shown on the contours below the red solid line. However it is subdominant above the red solid line. 
Notice that the produced DM tends to be warm and is intension with the Ly-$\a$ data for $m_a\lesssim \O(10)\KEV$. Therefore, the freeze-in region is disfavored. 
\paragraph{Constraints from topological defects/coherent oscillation} 
The topological defects or coherent oscillation contributes to the DM abundance depending on the nature of the DM mass. 
When $v_\F$ is sufficiently large, these contribution cannot be neglected. Therefore we do not consider the region above the blue dashed band. These production will be discussed in more details in later in this section.

\paragraph{Irrelevant constraints and consistency}
Before ending this section let us mention some constraints that are irrelevant and not shown in this figure. 
$a$ is in kinetic equilibrium with the thermal plasma, if the scattering of a process $a H\to a H$ is too fast. The scattering rate is given by
\beq
\G_{a H\to a H} \sim  \frac{\l_P^2}{4\pi^3 E_a} T^2,
\eeq
This form is justified when $E_a T \gtrsim m_\F^2$, where $E_a\sim m_\F ({T/T_{\rm crit}})$ is the energy of the produced NG boson energy. 
When $E_a T\lesssim m_\F^2$, it is much slower. 
Above the red line for the freeze-in, this process is always slower than the Hubble expansion. 

Generally, there is another contribution to the freeze-in production of $a$ from direct thermal scattering. 
The production rate via the portal coupling is given as 
\beq
\laq{HHaa}
\G_{ HH\to a a}\sim \frac{\lambda_P^2}{4\pi^3} T \text{~~~if~~~$T\gtrsim \max{[m_\F, m_h]}$} .
\eeq
When $T\lesssim m_\F$, this is suppressed since the NG boson-Higgs interaction comes from the higher dimensional term $\frac{\l_P}{2m_\F^2} |H|^2 (\partial a)^2$, which is generated by integrating out $s$. 
When $T\lesssim m_h$, it is suppressed by a Boltzmann factor.\footnote{Instead there are production processes via Higgs mixing, which is suppressed by the mixing angle.  } 
In the parameter region of focus, this production is subdominant compared with the $a$ production from the decay of the thermally produced $s$.

Since we assumed that $s$ never dominates the Universe to estimate the DM abundance, i.e. the remnant of $s$ does not dominate the Universe at $T_{\rm th}\lesssim T\lesssim T_{\rm crit}$, 
we need to check whether this is the case. 
In fact, this condition gives an upper bound of $v_\F$ which is much higher than the bound from topological defects/coherent oscillation. 
At the PT $s$ should oscillate, i.e.  $m_\F \gtrsim H_{\rm ubble}(T_{\rm crit})$, which is also satisfied in the shown region.

Lastly let us mention the fine-tuning on the SM Higgs boson mass. The dark Higgs field acquires a large vacuum expectation value which contributes to the SM Higgs boson mass via the portal coupling. 
It may be one of the sources of the fine-tuning problem of the SM Higgs mass, if this contribution is much larger than the electroweak scale.
Interestingly, in the viable parameter region the portal coupling contribution is negligible compared with the SM Higgs boson mass. 
In this sense, it may be viewed as a natural parameter region.

\subsection{Axion production via Peccei-Quinn symmetry breaking}

\label{sec:32}
Having discussed generic feature of the NG mode production at the symmetry breaking, now we look into more details of the case of axion DM.
Suppose that the global U(1) symmetry is explicitly broken by a small amount, which gives a potential for the NG mode, axion. The axion potential is assumed to be of the form
\beq
V_a=\L^4\(1-\cos{\(\frac{a}{f_a}\)}\).
\eeq
This can be either made if the ``Peccei-Quinn" (PQ) field $\F$~\cite{Peccei:1977hh,Peccei:1977ur,Weinberg:1977ma,Wilczek:1977pj}, which takes a role of dark Higgs field discussed so far, anomalously couples to some non-abelian gauge fields, which generate the axion potential due to non-perturbative dynamics or with some explicit breaking term like $\d {\cal L}\propto \F+\F^\dagger$. Here $f_a= \sqrt{2} v_\F/N_{\rm DW}$ with $N_{\rm DW}$ being the domain wall number, and we take  $N_{\rm DW}=1$ to evade the cosmological domain wall problem. 
In this case, domain walls are temporary formed at the onset of the oscillation of $a$, i.e. at $m_a\sim H$.  
However, each domain wall is bounded by a string. Soon after the domain wall formation the wall tension dominates the dynamics of the string-wall system and the domain walls collapse. 

In the axion model, we have two additional sources of the DM production other than that we have discussed so far, i.e. production at the symmetry breaking. 
One comes from the misalignment mechanism~\cite{Preskill:1982cy,Abbott:1982af,Dine:1982ah} i.e. from the axion coherent oscillation. The abundance is estimated as 
\beq
\Omega_a^{\rm mis} h^2\sim 10^{-3} \(\frac{f_a}{10^{10}\GEV}\)^{2} \sqrt\frac{m_a}{0.1\KEV},  \label{Omega_mis}
\eeq
where we have taken the misalignment angle $\theta_a=\pi/\sqrt{3},$ and we have assumed a temperature independent potential of $V_a.$ This contribution can explain the DM around the blue dashed line, above which it dominates  over our PT production. 
The other is the ALP radiation when the domain walls collapse~\cite{Yamaguchi:1998iv, Sikivie:2006ni, Kawasaki:2018bzv, Gorghetto:2018myk}. This contribution is more or less comparable to the misalignment one. 
Since there is a theoretical uncertainty on the numerical estimation of this contribution, we simply assume that these two contributions are the same order and we just use (\ref{Omega_mis}) as a representative one.
These contributions would dominate over our production mechanism at high $v_\F$ (much) above the blue dashed line in Fig.~\ref{fig:DM}.
This turns out to be subdominant due to the small decay constant in the region of interest (i.e. below the blue dashed line).

So far we have implicitly assumed that $m_a \AND f_a$ are independent. 
In the case of the QCD axion, which is well motivated from the viewpoint of strong CP problem in QCD,
the potential is generated via the non-perturbative dynamics of the  QCD and hence $m_a$ and $f_a$ are related. 
In this case we have $m_a f_a =\sqrt{\chi_0}$, where $\chi_0$ is the topological susceptibility, which we adopt $\chi_0 \approx (0.0756\GEV)^4$~\cite{Ballesteros:2016xej} (See also Refs.\,\cite{Berkowitz:2015aua,Bonati:2015vqz,Petreczky:2016vrs, Frison:2016vuc,Taniguchi:2016tjc}). 
The region compatible with this relation cannot be found in this figure because the DM is too heavy to be the axion. 
Strictly speaking, in the case of QCD axion, we need to take care of the existence of additional particles and topological defects, depending on the concrete UV completion model.
In the KSVZ model~\cite{Kim:1979if,Shifman:1979if}, we may have thermalized light PQ quarks in the symmetric phase.\footnote{We need a tiny mixing between the PQ fermion and the ordinary fermions to let the PQ fermions decay.}
There is no domain wall problem in the KSVZ model since $N_{\rm DW}=1$ with a minimal number of PQ quarks.
On the other hand, there is a domain wall problem in the DFSZ scenario~\cite{Dine:1981rt,Zhitnitsky:1980tq} in which $N_{\rm DW}=6$. To solve the problem we may introduce a tiny PQ breaking term in order to let the domain walls  collapse soon after the onset of the coherent oscillation of the axion. 
 Note that in the DFSZ model there is an additional Higgs doublet coupled to the PQ field, and hence the thermal potential discussed so far may be different, which may lead to a first order PT. We will come back to the possibility of the first order PT in Sec.~\ref{sec:4}.

\subsection{Hidden photon production via hidden $\U(1)$ breaking}
\label{sec:33}

Next we discuss the case of gauging the hidden $\U(1)$ symmetry in order to make the NG mode massive.
The Lagrangian of the hidden sector, including the Higgs-portal coupling, is given as 
\beq
\D {\cal L} \supset -\frac{1}{4}F_{\mu \nu} F^{\mu\nu}+ |D_\mu \F|^2 -V_\F(|\F|^2, |H|^2),
\eeq
where $F$ is the field strength of the Hidden photon, $g$ the gauge coupling, 
$D_\mu \F\equiv (\partial_\mu + i g A_\mu )\F $ is the covariant derivative of the dark Higgs.  
In this case, we can still calculate the (longitudinal component of the) hidden photon DM abundance from \eq{abundance} thanks to the equivalence theorem, by taking $m_a=m_A=\sqrt{2} g v_\F$ with $m_A$ being the mass of $A$. 
Since we are interested in the case of very small $g$, and since the interaction of the dark Higgs to the transverse gauge boson is suppressed by the coupling $g$, we can safely neglect the production of transverse mode. 

In principle we can write down the kinetic mixing term, $F_{\m\n}F_Y^{\m\n}$, between gauge fields of the SM $\U(1)_Y$ and hidden $\U(1)$. 
However this can be neglected if we take $g$ small enough or assume a charge conjugation symmetry in the hidden sector, $A\to -A, \F\to \F^\dagger$ to forbid the kinetic mixing.
In either case we do not need to care the thermal production of (transverse components of) the hidden photon via the gauge interaction. 

In the hidden photon model, there is an additional contribution to the hidden photon abundance from the cosmic string network formed during the symmetry breaking. As shown in Ref.~\cite{Long:2019lwl}, cosmic string loops emit (longitudinal component of) the hidden photon as far as the loop size is smaller than $m_A^{-1}$. This production is dominant above the blue dashed line, which is taken from Ref.~\cite{Long:2019lwl}. 
Another contribution may be from the inflationary period or (pre)heating~\cite{Graham:2015rva,Ema:2019yrd,Ahmed:2020fhc,Kolb:2020fwh}. This component, however, is sensitive to the inflation scale and the reheating dynamics, and is subdominant if the inflation scale is not very high and not shown here. 

An important difference between the axion case and hidden photon case is that cosmic string networks remain until present day in the latter case. The cosmic string tension is constrained by several observations. A robust constraint comes from the CMB observation, which indicates $v_\Phi\lesssim 2\times 10^{15}\,$GeV~\cite{Ade:2015xua}.
The cosmic string networks necessarily produce string loops in order to maintain the scaling solution, and string loops emit gravitational waves~~\cite{Vilenkin:2000jqa,Binetruy:2012ze}. In the present case, because of the smallness of the hidden photon mass, loops lose their energy dominantly through the emission of longitudinal vector boson if the loop size is smaller than $m_A^{-1}$ and through the gravitational waves if the loop size is larger~\cite{Long:2019lwl}. There are orders-of-magnitude uncertainties of the typical loop size, but for wide range of parameters the string loops contribute to the stochastic gravitational waves at the nano-frequency range, at which pulsar timing arrays have a good sensitivity. The recent NANOGrav result~\cite{Arzoumanian:2020vkk} gives an upper bound on the symmetry breaking scale as $v_\Phi\lesssim 5\times 10^{13}$\,GeV if the loop size is about one-tenth of the Hubble horizon scale, but it is relaxed as $v_\Phi\lesssim 10^{15}\,$GeV if the loop size is smaller~\cite{Blasi:2020mfx}. If the symmetry breaking scale is close to this upper bound, it is possible to explain the NANOGrav evidence of the gravitational waves.\footnote{
	Since the dominant contribution to the gravitational waves at the NANOGrav frequency range comes from loops that is going to decay at present, size of such loops is large enough to forbid the emission into the hidden photon. However, on the very high frequency range, at which laser interferometer gravitational wave detectors are sensitive, the signal may be greatly reduced due to the emission into the hidden photon. Such correlations between the low and high frequency gravitational wave signals may be a smoking-gun of this scenario.}

\section{Light dark matter from first order phase transition} \label{sec:4}

\begin{figure}[!t]
\begin{center}  
   \includegraphics[width=75mm]{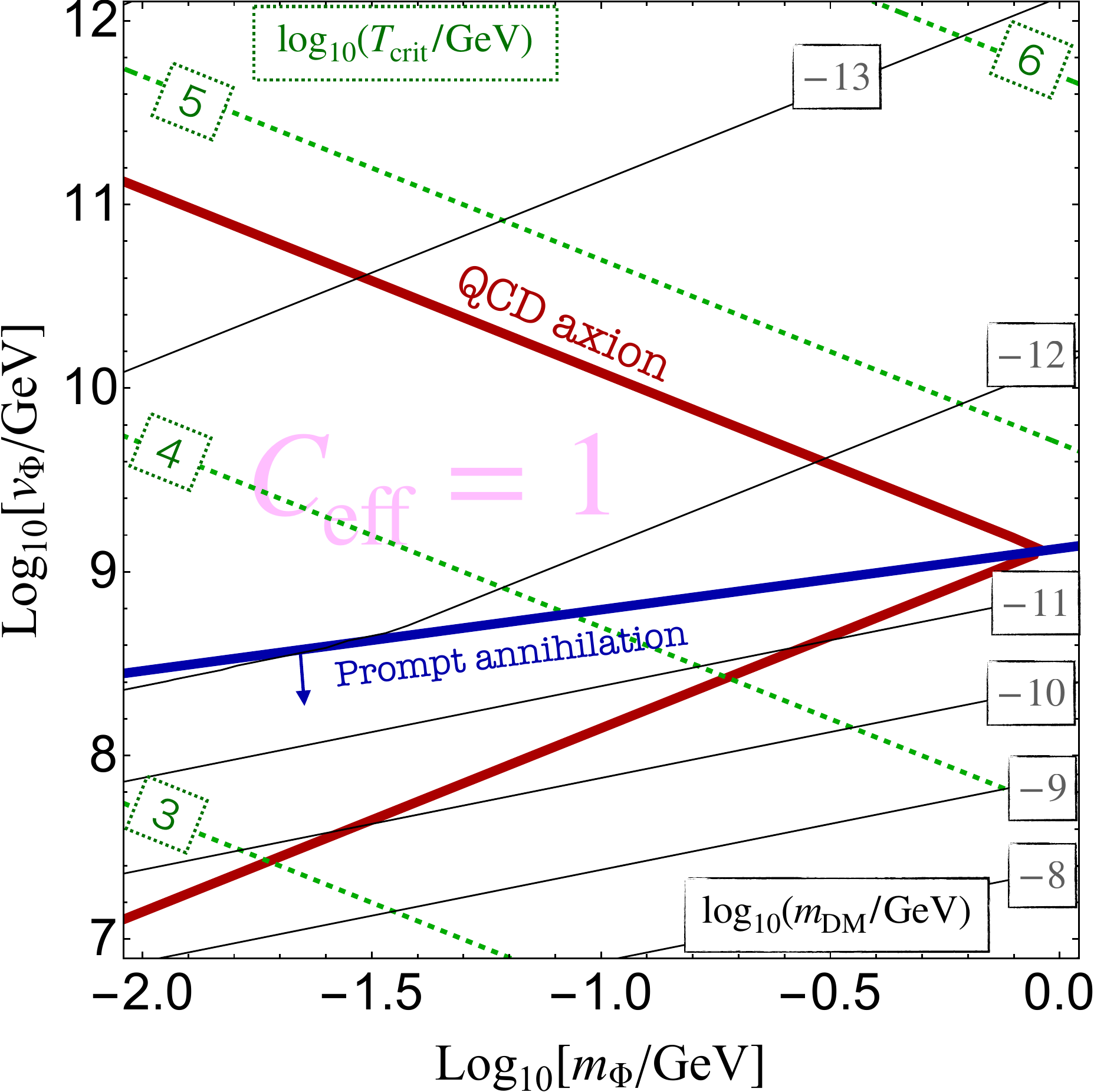}
      \includegraphics[width=75mm]{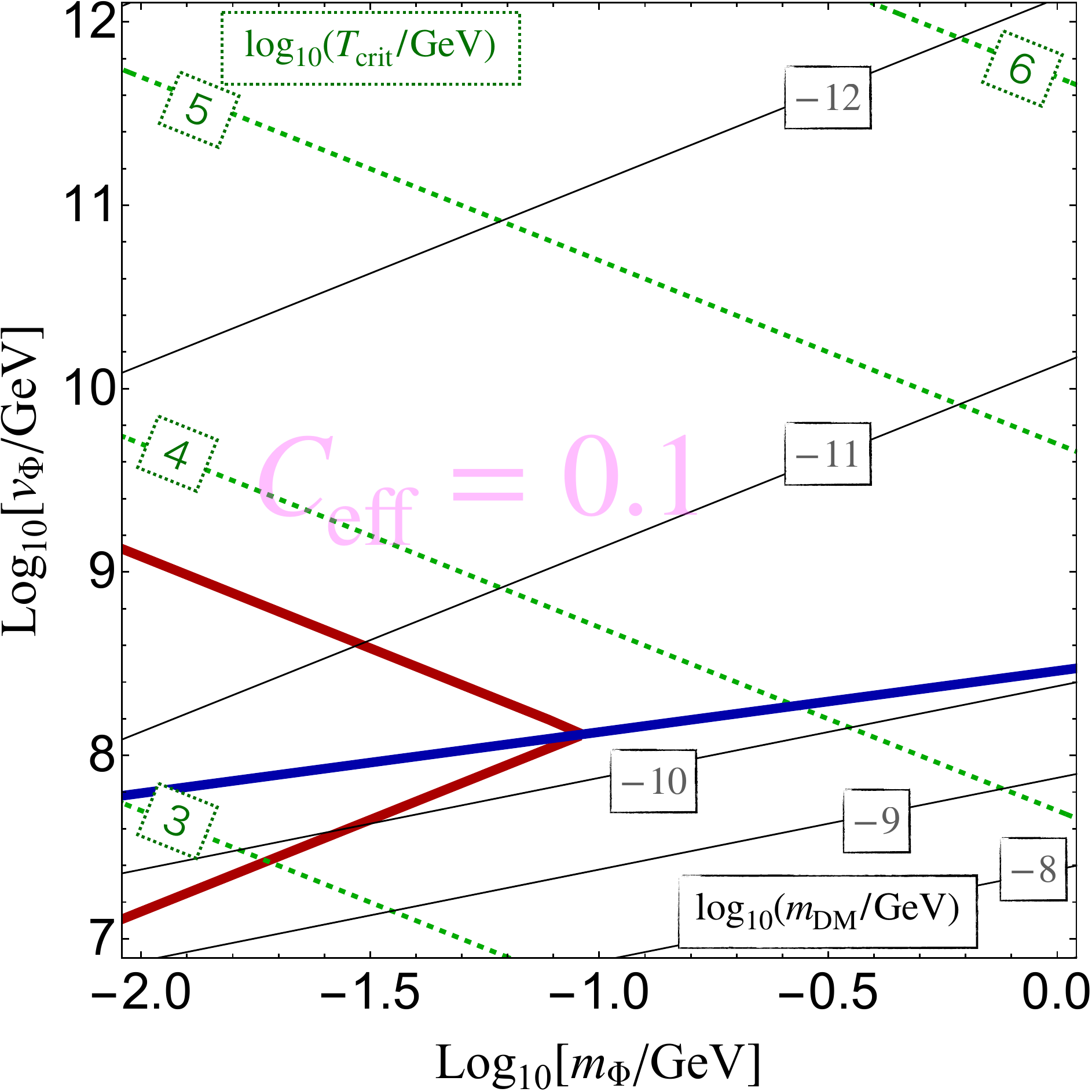}
      \end{center}
\caption{
The contours for the DM mass on $(m_\Phi, v_\F)$ plane in the first order phase transition with $C_{\rm eff}=1$ [left panel] and $0.1$ [right panel].  The contours of the temperature of PT, $T_{\rm crit}$, are also shown by green dotted lines. 
Below the blue solid line the produced DM soon annihilates. On the red line, the relation between the mass and decay constant is consistent with the QCD axion. 
} \label{fig:DM2}
\end{figure}

So far we have considered a simple setup of the second order phase transition of a hidden $\U(1)$ global or gauge symmetry in the early universe. 
This is true if the dark Higgs only has a portal coupling to the SM Higgs boson. On the other hand, the dark Higgs field may also have other couplings in general. 
In particular if the NG boson is the QCD axion, it should either coupled to heavy Higgs boson in the DFSZ model or PQ quarks in the KSVZ model. 
The inclusion of the new thermal and Coleman-Weinberg contributions to the potential may lead to a first order phase transition. 
In the first order PT, $s$ stays longer around the hilltop of the potential, then undergoes a tunneling and starts to oscillate. 
Therefore the suppression factor $C_{\rm eff}$ (\ref{Ceff}) tends to be close to one.

Strictly speaking, in such a Higgs potential that leads to a first order PT, the bubble wall may take away a fraction of the energy stored in the potential in the symmetric phase, 
like the well-known reheating problem in the old inflation. Then the Higgs oscillation amplitude in the broken phase should be suppressed according to energy conservation. 
This bubble wall expansion, however, gets a friction due to the pressure induced by the interactions the out-side thermal plasma and wall, and reach a terminal velocity~\cite{Dine:1992wr,Bodeker:2009qy,Bodeker:2017cim, Vanvlasselaer:2020niz} (see also Ref.~\cite{Hoeche:2020rsg}.)
in which case, we expect that the Higgs field in the broken phase exhibits a coherent oscillation.

In this section, let us simplify the discussion with the assumption that the PT takes place not too later than $T\sim T_{\rm crit},$ and the number density of the $s$ oscillation is given by \Eq{numd} with $C_{\rm eff}$ taken as a free parameter to take  account of the model-dependence and the uncertainty due to the bubble wall dynamics.  Moreover we neglect the effect on the NG boson production due to the bubble wall dynamics. 
By these assumptions, our previous discussions remain intact. 

In Fig.\ref{fig:DM2} we show the parameter region with $C_{\rm eff}\approx 1$ [left panel] and $C_{\rm eff} =0.1$ [right panel], by particularly focusing on the QCD axion range. 
Below the blue solid line, the number of the NG boson is produced too much initially, and the 
annihilation takes place promptly. The abundance is given by \Eq{anni}. 
We find that $0.01 \lesssim C_{\rm eff}\lesssim 1$, the QCD axion produced by the PT can explain the present DM abundance. 

In general, during the first order PT,  gravitational waves are produced via the bubble collisions or plasma sound wave~\cite{Kosowsky:1992rz,Kamionkowski:1993fg}. 
The typical frequency of the gravitational wave is determined by the bubble size, denoted by $\beta^{-1}$, at the collision. It depends on the details of the dark Higgs interactions, and it is estimated as
\beq
f_{\rm GW}\sim  0.2\,{\rm Hz}\(\frac{T_{\rm reh}}{1\TEV}\) \left( \frac{\beta/H_{\rm ubble}(T_{\rm crit})}{10^3} \right),
\eeq
where $T_{\rm reh}$ is the thermal temperature at the completion of the PT. 
In our scenario, $T_{\rm reh}\sim T_{\rm crit}$. 
In Fig.\ref{fig:DM2} contours of the critical temperature $\sim T_{\rm reh}$ are shown by the green dotted lines. 
This does not depend on $C_{\rm eff}$. Interestingly, when the QCD axion DM is successfully produced during the PT, we obtain 
$ 1\,{\rm mHz} \lesssim f_{\rm GW} \lesssim 10\,{\rm Hz}$ taking account of the model dependence of $\beta/H_{\rm ubble} \sim 10$--$10^3$, which may be within the sensitive range of LISA\cite{Audley:2017drz} and DECIGO~\cite{Kawamura:2020pcg}.

\section{Conclusions and discussion}

\label{sec:dis}

In this paper we have proposed the hidden photon or axion-like particle DM production via continuous symmetry breaking with a dark Higgs, taking account of the interaction between the dark Higgs and the SM Higgs. 
It is a minimum setup that accounts for the dark global or gauged U(1) symmetry breaking.
Even in this simple setup, the dark Higgs dynamics and its consequence for the DM production are complicated partly due to thermal effects. 
We found parameter regions that are consistent with present DM abundance. In our scenario the DM mass can be as light as 1\,${\rm eV}$.
The light DM may be warm and can be tested in the future observations of the 21cm line~\cite{Sitwell:2013fpa}. 
On the other hand, given a setup that the PT is the first order, the DM can be much lighter and there is a possibility that the QCD axion produced by the dark Higgs dynamics takes a role of DM. In this case, the gravitational waves from the PT may be tested in the future. 

Some additional comments are in order.
In the main part we have considered the case where the dark Higgs does not dominate the energy density of the Universe before the PT.
An interesting alternative possibility may be that $s$ is the inflaton, which means $s$ dominates the Universe and must reheat the Universe later. 
Let us suppose that $\U(1)$ is gauged and $g$ is chosen so that $H_{\rm inf}^2/M_{\rm pl}^2\sim V_\F(0)/M^4_{\rm pl} \sim g^2.$ Then the curvature at the hilltop of the potential may satisfy \beq g^2 M_{\rm pl}^2 \geq V_\F''(>0),\eeq according to the weak gravity conjecture (WGC)~\cite{ArkaniHamed:2006dz, Cheung:2014vva}.\footnote{We are not sure if the WGC can work in a false vacuum. However, this issue also exists in the original paper explaining the hierarchy problem of the SM since the electroweak vacuum is essentially false vacuum in the SM~\cite{Cheung:2014vva}.  }  
The least tuned region saturates the WGC~\cite{Cheung:2014vva,Yin:2020dfn}, $g^2M^2_{\rm pl}= V_\F''.$
Thus we may have a local tiny minimum at the potential maximum. 
An old inflation takes place there (for the e-folds, see e.g. Refs.~\cite{Kitajima:2019ibn,Matsui:2020wfx}) and later $s$ tunnels through the potential barrier.
After the tunneling, still the curvature of the potential may be suppressed enough and then the quartic hilltop inflation happens there~\cite{Izawa:1996dv,Asaka:1999jb,Senoguz:2004ky} if the quartic coupling is negative. 
Note that the WGC required the potential to be flat and the slow-roll condition is satisfied. 
The potential can have a minimum stabilized by the quartic term and a higher dimensional term~\cite{Nakayama:2011ri, Nakayama:2012dw, King:2017nbl,  Antusch:2018zvu, Guth:2018hsa, Matsui:2020wfx}.\footnote{The inflaton, $s$, can also be stabilized by two or more higher dimensional terms. In such a case the VEV and the mass of $s$ are the typical scales of the higher dimensional terms~\cite{Jaeckel:2020oet}. }
Soon after the slow-roll inflation ends, the hidden photon DM is produced via the parametric resonance as we have discussed in the main part. 
However, as we have also mentioned, the $s$-condensate may not completely disappear and remain slightly. This may dominate the Universe again at the later stage, and then its decay reheats the Universe again. 
The detail of the reheating is complicated due to thermal corrections and we leave it for our future study. 
In any case, irreverent to the detailed thermal history, an unique prediction is the relation between $m_A$ and $H_{\rm inf}$,
\beq
m_{A}= \sqrt{2}g v_\F\sim  \frac{v_\F H_{\rm inf}}{M_{\rm pl}}.
\eeq
In particular, if the higher dimensional terms are suppressed by $M_{\rm pl}$, $v_\F\sim 10^{12}\GEV$ and $H_{\rm inf}\sim 1\GEV$~\cite{Guth:2018hsa}. 
This predicts $m_A\sim \KEV \AND g\sim 10^{-18}.$

\appendix

\section{Parametric resonance in thermal environment}
\label{ap:1}
Here let us discuss whether significant particle production via parametric resonance may occur in a thermal environment. 
This is important since in a large parameter region the resonance parameter for the $s$-$H$ system is larger than unity.

As a toy model, we consider 
\beq
{\cal L}\supset -(g |H|^2+m_\F^2)\frac{s^2}{2},
\eeq
where $g$ is a portal coupling, $H$ is a scalar complex field and is massless at the vacuum $s=0$. In this Appendix we neglect the expansion of the Universe for simplicity. Later we will discuss the case where $H$ is the SM Higgs, but for a while we keep $H$ just as a general complex scalar field.
As is well known, particle production of $H$ happens due to the $s$ coherent oscillation. 
When $H$ is coupled to thermal bath, this process becomes more involved. 

Let us first consider the case $H$ does not interact with any other particle and it is only produced by $s$.  
Then a broad parametric resonance occur if $q\sim g s^2_{\rm amp}/m_\F^2\gg 1$. The number density of $H$, $n_H$, increases exponentially. 
We emphasize that $s$ is now wave-like, and we cannot describe the evolution from perturbation theory for particles. 
For instance one can easily find the $s^n\to H H$ process has a rate proportional to $q^{n} m_\F$, and the particle picture is highly non-perturbative. We can describe the evolution of $s$ by solving the equation of motion 
 (in the 1PI effective theory). 
Following~\cite{Kofman:1997yn}, we can write down the equation of motion (EOM) of $s$ as
\beq
\ddot{s}= -m_\F^2 s -g  \vev{|H|^2} s+\cdots
\label{s_eom_app}
\eeq
where $\vev{H^2}\equiv \int {\frac{d^3k}{(2\pi)^3} |H_k|^2},$
and $\cdots$ represents terms with higher order in $s$, $s_k$. These neglected terms would be important when resonance lasts long enough and the processes known as re-scattering would occur, which, however, is not our focus. 
We emphasize that this EOM should include all the effects (of non-perturbative series in the particle picture) which only involve the $s$ zero modes. 

By using $\vev{|H|^2}\sim 2n_H/m_H(s)$, 
with 
$m_H[s]=\sqrt{g} |s|$,
one obtains \beq g  \vev{|H|^2} s = \sqrt{g} n_H\,{\rm sign}{(s)}.\eeq
With this, the EOM looks like that  $s$ moves in an effective potential of $V_{\rm eff} \sim  m_\F^2 s^2/2+\sqrt{g} n_H |s|.$
If $n_H$ increases $s_{\rm amp}$ should decreases so that $V_{\rm eff}[s_{\rm amp}]$ is kept. 
Since $n_H$ increases exponentially due to the parametric resonance, the amplitude of $s$ is decreased. \\

Now let us consider the parametric resonance in thermal environment where $H$ is thermalized with a temperature $T$ and suppose that the $s$ oscillation time scale is much longer than the thermalization time scale: $m_\Phi \ll T$ so that the $s$ oscillation is nearly adiabatic with respect to the thermal bath. 
There may be a coupling of $H$ to other fields like gauge bosons, with typical coupling of $g' \sim 1$, which may induce the $H$'s thermal mass of $\sim g' T$. Then the total mass of $H$ is expressed as 
\beq
m^2_H[s]\sim \sqrt{ g'^2 T^2+ g s^2}. 
\eeq
The energy density of $H$ is expressed as 
\beq
\rho_H \sim T^4.
\eeq
Since we assume that $\rho_s$  is smaller than $\rho_H$, $\rho_H$ is kept almost constant during the oscillation of $s$. 
This means that the two point function, satisfying $\vev{|H|^2 E_H^2} \sim \rho_H,$ is bounded by
\beq
T^2 \lesssim \vev{|H|^2} \lesssim T^4/m^2_H[s] \sim T^2 g'^{-2} + g g'^{-4} s^2+\O(s^4/T^2).
\eeq
In the left hand side we divide $\rho_H$ by maximal Higgs energy $T^2$ while in the right hand side we divide it by the minimal one, $m_H^2.$ 
This implies that $\vev{|H|^2}$ cannot change much for $g'\sim 1$ and hence  $s_{\rm amp}$ does not decrease much. 

Let us estimate the dissipation rate of $s$ in this setup following the arguments in Refs.~\cite{Berera:1995ie,Berera:1998gx,Yokoyama:1998ju}.
We introduce a time scale $\Delta t(k)$, which represents a typical time scale for the thermal distribution of $H$, i.e., the time scale for a distribution function $n_k\equiv |H_k|^2 w_k$ reaches to the equilibrium distribution $n_k^{\rm eq}=(-1+\exp{(-\sqrt{k^2+g s^2} /T)})^{-1}.
$ with $w_k= \sqrt{k^2+g^2 s^2}$.
 The $k$ dependence of $\D t[k]\propto g'^4$ is model dependent and we here 
 assume that $\D t$ decreases fast enough if $k$ is smaller than $T$, i.e. the scatterings  of IR modes are efficient. 
Since $s$ is time varying and the thermalization time scale is finite, 
$n_k$ at the time $t$ exhibits the equilibrium distribution at slightly earlier time $t-\D t[k]$,  
\beq
n_k(t)\sim n^{\rm eq}_{\rm k}(t-\D t[k]).
\eeq
Thus we can estimate
\begin{align}
\laq{defeq}
\vev{|H|^2}&\sim \int{\frac{d^3 k}{(2\pi)^3 w_k}} \left(n^{\rm eq}_k- \D t[k] \frac{d}{dt} n^{\rm eq}_k\right) \\
&=\frac{1}{12} T^2-\frac{\sqrt{g} s}{2\pi^2 T}+\O(s^2/T^2) +C \D t[T]  \frac{g \dot{s} s}{2\pi^2} +\O(1/T^3),\laq{dev}
\end{align}
with $C$ being an $\O(1)$ numerical coefficient.
We note that $s$ is time-dependent and the time derivative in \Eq{dev} is non-vanishing. 
 \Eq{dev} represents the deviation from the thermal equilibrium, 
 and we approximated the dominant contribution from the integrant around $k\sim T$ since when $k\ll T$ ($k\gg T$) it is suppressed by $\D t$ (Boltzmann suppressed).
By inserting \Eq{dev} into the EOM (\ref{s_eom_app}) and multiply both sides by $2\dot{s}/m_\F$, we obtain the evolution equation for the number density $n_s$.
Then it is found that the term proportional to $\dot s$ in \Eq{dev} leads to the effective friction of $s$ and leads to the dissipation of $s$ energy density.  
We note that by taking a time average (over a few $2\pi /m_\F$), terms without time derivatives in the EOM (\ref{s_eom_app}) are cancelled out. We then arrive at
\beq
\dot{n}_s\sim - C \D t [T] \frac{2g^2 n_s^2}{m_\F\pi^2} .
\eeq
 This is smaller than the na\"{i}ve estimation of $s$ annihilation contribution $\sim n_s^2\times \s_{ss}\sim g^2 n_s^2/(4\pi m_\F^2)$ 
 since $\D t[T] m_\F \ll1.$
  By counting the number of $s$ one may identify the process corresponding to the particle picture (at least in $q\ll1$ limit). For example, $s^{2n} \dot{s}$ term in the EOM should correspond to the scattering of $n$ zero modes of $s$. As we can see  it is suppressed by $(g s^2/ T^2)^n$.

For the symmetry breaking system discussed in the main part of this paper, we similarly obtain
\beq
\dot{n}_s\sim   - \lambda_P^2 v_\F^2 \D t[T]\frac{n_s}{\pi^2 }.  
\eeq
The leading term, by noting $\D t[T]\sim (y_t^2 T)^{-1}$, corresponds to the dissipation term \eq{diss}.

\section{Dissipation of (would-be) NG boson condensate}

\label{sec:dissipation}

To discuss the dissipation of the produced NG boson or would-be NG boson, whose momentum is much smaller than the temperature and the occupation number is extremely large, we may also apply a similar method to the case of dissipation of $s$ given in App.~\ref{ap:1}.\footnote{
	The dissipation of QCD axion, which is coupled to the gluon though the anomaly, has been discussed in Ref.~\cite{Moroi:2014mqa}. In our present model, we do not necessarily assume such interactions and the dominant source of axion dissipation comes from the interaction with the SM Higgs (\ref{axion_Higgs}).
}
The Lagrangian under consideration is 
\beq
{\cal L} \supset \frac{1}{2}\partial_\mu a \partial^\mu a -\frac{\l_P }{2m_\F^2}(\partial_\mu a)^2 |H|^2.
\label{axion_Higgs}
\eeq
By assuming spherical symmetric distribution of $a_{\vec{k}}$ and assuming that only $|\vec{k}|=k$ modes dominate, the Hartree approximation reads 
\beq
\ddot{a}_k=-k^2 a_k+\l_P\frac{ 1 }{m_\F^2} \dot{a}_k  \vev{\dt|H|^2}+\cdots
\label{a_eom_app}
\eeq
where $\cdots$ includes $a_{k'\neq k}$ modes or higher order in $1/m_\F^2$. 
We neglect the other modes again due to the small occupation (note that in our scenario $q_a m_\F\sim m_\F \gg H_{\rm ubble}$, which means that the NG bosons are soon produced and the spectrum is nearly monochromatic). 
Then we obtain 
\beq
\laq{dH2}
\vev{\dt |H|^2}\sim  \int{\frac{d^3 k}{(2\pi)^3 w_k} \(\dt n_k^{\rm eq}-\D t[k] \frac{d^2}{dt^2} n^{\rm eq}_k \)} .
\eeq
Assuming $a_k\simeq a_k^{\rm amp}[t]\cos[k t]$, we obtain the equation for the evolution of number density $n_{a_k}$ from the equation of motion (\ref{a_eom_app}) as\footnote{
	One might think that the presence of NG boson particles does not affect the Higgs dispersion relation since $\left<(\partial_\mu a)^2\right>=0$ for massless NG boson and the interaction is of the form (\ref{axion_Higgs}). However, since the typical NG boson oscillation time scale $k^{-1}\sim (m_\Phi/2)^{-1}$ is much longer than the Higgs thermalization time scale, one should be careful about the time dependence of $\left<|H|^2\right>$ before taking time average.
}
\begin{align}
\dot{n}_{a_k}\approx  \frac{1}{2k t_{\rm ave}}\int_{t-t_{\rm ave}/2}^{t+t_{\rm ave}/2}{dt \frac{\l_P  }{m_\F^2} \dot{a}^2_k \vev{\dt |H|^2}} ,
\end{align}
with $n_{a_k}[t]=a_k^{\rm amp}[t]^2 k/2,$ $t_{\rm ave}$ is a time scale much longer than $k$ but so short that $a_k^{\rm amp}$ can be taken as constant. 
We note the contribution from the  first term of \eq{dH2} is negligible with large enough $t_{\rm ave}$ since it includes terms of even number of $\dot{a}_k$. 
In this case, the integral consists only total derivatives by using $\ddot{a}_k[t]\approx -k^2 a_k[t].$
The non-vanishing contribution comes from the second term of \eq{dH2}, which has a leading contribution of 
\beq
\vev{\frac{d}{dt}|H|^2}\sim \int{\frac{d^3 k'}{(2\pi)^3 w_k'}\D t[k'] \frac{\lambda_P k^2 }{m_\F^2 T k'} (k^2 a_k^2-\dot{a}_k^2)\exp{(-k'/T)}+\O(\lambda_P^2/m_\F^4)}.
\eeq
By assuming again that $\D t[k]$ is larger at larger $t$ we obtain the integral dominates at around $k'\sim T,$ 
and 
\beq
\vev{\frac{d}{dt}|H|^2}\sim  C \D t[T]\frac{1}{4\pi^2} \frac{\l_P k^2}{m_\F^2 } (k^2a^2_k-\dot{a}_k^2).
\eeq 
Substituting this expression into the equation of motion (\ref{a_eom_app}) and performing the $t$ integral by using the explicit form of $a_k[t]$ and $n_{a_k}$, we arrive at   
\beq
\dot{n}_{a_k}\sim - C \D t [T] \frac{\lambda_P^2  k^3 n_{a_k}^2}{ 8 \pi^2 m_\F^4} .
\eeq
This gives the dominant contribution to the $k$ modes scattering of the NG bosons.
This is equivalent to have a dissipation rate of 
\beq
\G_{\rm dis}^{(a)}\sim\frac{\lambda_P^2  k^2 n_{a_k}}{8 \pi^2 m_\F^4 } \times \frac{k}{y_t^2 T},
\label{axion_diss}
\eeq
where we have used $\D t [T]\sim \(y_t^2 T\)^{-1}$.

The same result is also obtained from a diagrammatic approach.
The optical theorem tells us that the annihilation cross section of the NG bosons is given by $\left<\sigma v\right>_{aa} \sim (k^0)^{-2}\,{\rm Im} \mathcal M(aa\to aa)$, where $k^0$ denotes the total incoming NG boson energy and $\mathcal M(aa\to aa)$ the amplitude. Since the SM Higgs obtains large thermal mass, we need to take account of its thermal width for the Higgs propagating in the loop. At the one-loop level, it is evaluated as~\cite{Mukaida:2012qn}
\begin{align}
	{\rm Im} \mathcal M(aa\to aa) \sim \lambda_P^2 \int d^4q \left[f_{\rm B}(q^0)-f_{\rm B}(q^0-k^0) \right]\rho(q^0)\rho(q^0-k^0),
\end{align}
where $f_{\rm B}(q^0) = (e^{q^0/T}-1)^{-1}$ is the Bose-Einstein distribution and we take the Breit-Wigner form for the spectral density
\begin{align}
	\rho(q^0) = \frac{q^0 \Gamma_{\rm th}}{[(q^0)^2-(\vec q^2 + m_H^2)]^2 + (q^0 \Gamma_{\rm th})^2},
\end{align}
with $\Gamma_{\rm th}$ being the thermal width of the SM Higgs, which is expected to be $(\Delta t[T])^{-1}$.
Assuming $\Gamma_{\rm th}\gg m_\Phi (\sim k^0)$, we obtain the dissipation rate for the NG boson through $\G_{\rm dis}^{(a)} \sim \left<\sigma v\right>_{aa} n_{a_k}$ and the result is the same as (\ref{axion_diss}).

\section*{Acknowledgements}

This work was supported by JSPS KAKENHI Grant (Nos. JP19J13812 [KN], 18K03609 [KN], 17H06359 [KN], 16H06490 [WY], 19H05810 [WY] and 20H05851 [WY].)
WY would like to thank KEK for the kind hospitality when part of this work is done.

\end{document}